\documentclass[a4paper,11pt]{article}
\usepackage[english]{babel}
\usepackage[utf8]{inputenc}
\usepackage{braket}
\usepackage{physics}
\usepackage{color}
\usepackage{amssymb}
\usepackage{amsmath}
\usepackage{graphicx}
\usepackage{epstopdf}
\usepackage{subcaption}
\usepackage{comment}
\usepackage{mathtools}
\usepackage{ragged2e}
\usepackage{hyperref}
\hypersetup{
	colorlinks=true,
	linkcolor=blue,
	filecolor=cyan, 
	urlcolor=blue,
	citecolor=red,
} 
\usepackage[titletoc,toc,title]{appendix}
\numberwithin{equation}{section}
\usepackage{cleveref}
\usepackage[margin=2.5cm]{geometry}
\usepackage{cite}
\usepackage{epsfig}
\usepackage{float}
\usepackage{cancel}
\usepackage{amsfonts}
\usepackage{enumitem}
\usepackage[font={footnotesize,it}]{caption}
\usepackage{authblk}
\begin{document}
	
	\title{Balanced Partial Entanglement in Flat Holography}
	
	\author[1]{Debarshi Basu\thanks{\noindent E-mail:~ {\tt debarshi@iitk.ac.in}}}

	\affil[1]{
		Department of Physics\\
		
		Indian Institute of Technology\\ 
		
		Kanpur 208 016, India
	}
	
	\date{}
	
	\maketitle
	
	\thispagestyle{empty}
	
	\begin{abstract}
		
		\noindent
		We advance a construction for the balanced partial entanglement entropy (BPE) for bipartite mixed states in a class of $(1+1)$-dimensional Galilean conformal field theories dual to Einstein gravity and topologically massive gravity in asymptotically flat spacetimes. We compare our results with the entanglement wedge cross section (EWCS) in the dual geometries and find exact agreement. Furthermore, utilizing a geometric picture, we demonstrate the holographic duality between the EWCS and the BPE in the context of flat holography. 
		
		\justify

	\end{abstract}
	
	\clearpage
	
	\tableofcontents
	
	\clearpage

\section{Introduction}
Quantum entanglement in extended many body systems plays a significant role in diverse areas of physics ranging from condensed matter systems to black holes in quantum gravity. In the recent past, extensive studies on various entanglement measures have revealed fundamental insights. For bipartite pure states the nature of quantum entanglement is characterized by the entanglement entropy which has been studied extensively in quantum field theories. However, the entanglement entropy fails to be a viable measure of entanglement for mixed states due to the appearance of irrelevant correlations. The characterization of entanglement in mixed states is a subtle and complex issue and to address this several measures like the entanglement of purification \cite{2002}, the logarithmic negativity \cite{Vidal:2002zz}, the reflected entropy \cite{Dutta:2019gen} and the mutual information have been introduced. In this context, the entanglement contour \cite{2014} or the partial entanglement entropy (PEE) \cite{Wen:2018whg,Wen:2019iyq,Wen:2020ech} serves as a local measure of quantum entanglement and may be utilized to characterize the fine structure of entanglement and correlation in mixed states.

In the context of the AdS/CFT correspondence, a holographic characterization of the entanglement entropy for a subsystem in CFT$_d$s dual to static AdS$_{d+1}$ geometries was forwarded in the seminal work of Ryu and Takayanagi \cite{Ryu:2006bv} which involves the area of a bulk codimension two static minimal surface (RT surface) homologous to the subsystem. A covariant generalization of the RT prescription was proposed in \cite{Hubeny:2007xt} and in a series of subsequent works \cite{Casini:2011kv,Lewkowycz:2013nqa,Dong:2016hjy} these proposals were proved. Following these developments the holographic characterization of the structure of mixed state entanglement in dual field theories has gained a lot of attention recently. In this context, for mixed states in holographic theories the entanglement wedge cross-section (EWCS) \cite{Takayanagi:2017knl,Nguyen:2017yqw} serves as a natural candidate to characterize the entanglement and correlation structure. Several information theoretic quantities including the reflected entropy \cite{Dutta:2019gen}, the odd entropy \cite{Tamaoka:2018ned} and the entanglement negativity \cite{Kudler-Flam:2018qjo,Kusuki:2019zsp,KumarBasak:2020eia,KumarBasak:2021lwm} have been proposed as putative holographic duals of the EWCS. In particular, the holographic duality between the reflected entropy and the EWCS was established in \cite{Dutta:2019gen}, using the gravitational path integral techniques developed in \cite{Lewkowycz:2013nqa}. Furthermore, in \cite{Wen:2021qgx}, it was demonstrated that in the context of generic entanglement purification of a bipartite mixed state $\rho_{AB}$ through the inclusion of an auxiliary system $A'B'$, the EWCS is directly related to certain PEEs satisfying specific \textit{balance conditions}. This quantity, termed the balanced partial entanglement (BPE) \cite{Wen:2021qgx,Camargo:2022mme}, was also shown to be equal to half of the reflected entropy for the case of canonical purification thereby elevating its status to a more generic correlation measure for mixed states. Another quantity of interest is the crossing PEE between $A$ and $B'$ under the balance conditions, which was shown in \cite{Wen:2021qgx} to be the natural generalization of the \textit{Markov gap} \cite{Hayden:2021gno} which quantifies how precisely a given tripartite mixed state can be recovered given the information of any two parties. It was demonstrated in \cite{Wen:2021qgx} that for bipartite states in holographic CFT$_2$s, the crossing PEE is an universal constant with the geometric interpretation in terms of the number of boundaries of the EWCS similar to the Markov gap \cite{Hayden:2021gno}.

In a separate context, a class of $(1+1)$-dimensional non-relativistic field theories with Galilean conformal symmetry (GCFT$_2$) \cite{GCA,Bagchi:2009pe,Correlation,Bagchi2015EntanglementEI,Basu_2016, Log,Alishahiha:2009np,Nishida:2007pj,Bagchi:2010vw,Hosseini:2015uba,Lukierski:2005xy,PhysRevD.5.377,0264-9381-10-11-006,Martelli:2009uc,Duval:2009vt} have been proposed to have $(2+1)$-dimensional asymptotically flat geometries as their gravitational duals \cite{Bagchi2010CorrespondenceBA,Barnich2010AspectsOT}. The entanglement structure of bipartite pure and mixed states in such GCFT$_2$s were investigated utilizing suitable replica techniques in a series of communications \cite{Bagchi2015EntanglementEI,Basu_2016,Malvimat:2018izs,Basak:2022cjs,Basak:2022gcv}. In the framework of flat space holography \cite{Bagchi2010CorrespondenceBA,Barnich2010AspectsOT}, the bulk dual spacetimes for such GCFT$_2$s may involve pure Einstein gravity or topologically massive gravity \cite{Deser1982ThreeDimensionalMG,Deser1982TopologicallyMG,Jiang:2017ecm,Skenderis_2009,Hijano:2017eii,Castro_2014} and the asymptotic symmetries of these $3d$ geometries are described by the Galilean conformal algebra (GCA$_2$). A holographic characterization of the entanglement entropy in these flat holographic scenarios was established in \cite{Jiang:2017ecm,Hijano:2017eii} through a novel geometric construction similar to the HRT prescription \cite{Hubeny:2007xt}. Subsequently, a covariant geometric picture of the entanglement wedge cross-section for bipartite mixed states in holographic GCFT$_2$s was developed in \cite{Basu:2021awn} in order to investigate a relation between the EWCS and the holographic entanglement negativity in such non-relativistic theories\footnote{For relevant discussions, see \cite{Basu:2021axf,Setare:2021ryi}.}. Remarkably, recent field theoretic replica calculations for the reflected entropy for various bipartite states in such GCFT$_2$s \cite{Basak:2022cjs,Setare:2022uid} match exactly with the EWCS obtained in \cite{Basu:2021awn} in the large central charge limit.  

As mentioned earlier, the BPE is a natural candidate for the characterization of mixed state entanglement and their purification. Moreover, the BPE may be defined for generic theories and therefore its correspondence with the EWCS or the reflected entropy should hold beyond the framework of AdS/CFT. The authors in \cite{Camargo:2022mme} recently investigated the BPE in holographic Bondi-Metzner-Sachs (BMS) field theories \cite{Barnich2006ClassicalCE,Barnich2010AspectsOT,Bagchi2012BMSGCART} with Einstein gravity duals. In this article, we establish the holographic duality between the BPE in large central charge GCFT$_2$s and the minimal EWCS in asymptotically flat bulk $3d$ geometries involving pure Einstein gravity as well as topologically massive gravity. Remarkably our results match exactly with the minimal EWCS computed in \cite{Basu:2021awn} for each of these bipartite states. We also compute the crossing PEEs for bipartite mixed state configurations involving two adjacent intervals and find that at the balance points they vanish for pure Einstein gravity in the bulk. This indicates that there may be a perfect Markov recovery from mixed states in the dual GCFT$_2$s \cite{Hayden:2021gno,Basu:2021awn,Camargo:2022mme}. On the other hand, for TMG in the bulk geometries the PEE turns out to be a universal constant with the geometric interpretation in terms of the number of boundaries of the EWCS similar to the AdS/CFT scenario \cite{Hayden:2021gno}. Furthermore, we provide a geometric picture of the BPE in these flat holographic setups and demonstrate its equivalence with the EWCS constructed in \cite{Basu:2021awn}.

The rest of this article is organized as follows. In \cref{sec:review} we review  the basic definitions of the entanglement contour and the BPE. In \cref{sec:BPE-flat}, following a brief review of the various entanglement measures in flat holographic settings, we compute the BPE for various bipartite mixed states in GCFT$_2$s with Einstein gravity and TMG duals. We also provide the crossing PEEs for the mixed states of two adjacent intervals and comment on the Markov recovery process. Finally, in \cref{sec:summary}, we present a summary of our work and discuss some possible future directions.

\section{Review: Balanced partial entanglement}
\label{sec:review}
\subsection{The partial entanglement entropy}
The partial entanglement entropy (PEE) is a local measure of the entanglement structure which measures the contribution to the entanglement entropy of a subsystem $A$ from a subset $A_i\in A$ as follows \cite{Wen:2018whg,Han:2019scu,Wen:2019iyq,Wen:2020ech}
\begin{align}
	s_A(A_i)\equiv\mathcal{I}(A_i,A^c)=\int_{A_i}s(x)\,dx\,,
\end{align}
where $s_A(A_i)$ is the PEE and $s(x)$ is the fine-grained entanglement contour function discussed in \cite{2014,Wen:2018whg,Han:2019scu,Wen:2019iyq,Wen:2020ech}. The PEE quantifies a kind of mutual correlation between the subset $A_i$ and the complement $A^c$ of $A$ and satisfies various physical properties such as additivity, positivity, permutation symmetry and unitary invariance \cite{Wen:2018whg,Han:2019scu,Wen:2019iyq,Wen:2020ech}. Furthermore, $\mathcal{I}(A,B)$ is bounded by the minimum of the entropies for each subsystem and satisfies the normalization condition
\begin{align}
	\mathcal{I}(A,B)\big|_{B\to A^c}=S(A)\,.
\end{align}
Recently, a suitable proposal to compute the PEE in two-dimensional quantum field theories was forwarded in \cite{Wen:2018whg,Wen:2020ech,Wen:2019iyq,Kudler-Flam:2019oru} in terms of an appropriate linear combination of the entanglement entropies of various subsystems\footnote{See \cite{Lin:2021hqs} for a derivation of the PEE proposal from a locking bit thread configuration.}. For a subsystem $A$ partitioned into three non-overlapping subsets $A=\alpha_L\cup \alpha\cup \alpha_R$, the above proposal claims that the PEE $s_A(\alpha)$ is given as
\begin{align}
	s_A(\alpha)=\frac{1}{2}\left(S(\alpha_L\cup \alpha)+S(\alpha\cup\alpha_R)-S(\alpha_L)-S(\alpha_R)\right)\,.\label{PEE-proposal}
\end{align}
In the above partition, $\alpha_{L(R)}$ denotes the degrees of freedom in $A$ that are on the left (right) of the subset $\alpha$ and therefore a natural ordering of the subsets is assumed. Interestingly, this alternating linear combination (ALC) proposal satisfies all the properties of the PEE \cite{Wen:2018whg,Han:2019scu,Wen:2019iyq,Wen:2020ech}.

\subsection{Balanced partial entanglement}
Having reviewed the salient features of the PEE proposal, we now provide the definition of the balanced partial entanglement (BPE) \cite{Wen:2021qgx,Camargo:2022mme} in the following. To begin with, one considers a bipartite mixed state on $\mathcal{H}_A\otimes \mathcal{H}_B$ described by the density matrix $\rho_{AB}$. A pure state $|\psi\rangle$ may be obtained from $\rho_{AB}$ by introducing an auxiliary purifier system $A'B'$, such that 
\begin{align}
 \text{Tr}_{A'B'}|\psi\rangle\langle\psi|=\rho_{AB}~~,~~|\psi\rangle\in \mathcal{H}_A\otimes \mathcal{H}_B\otimes\mathcal{H}_{A'}\otimes \mathcal{H}_{B'}\,.
\end{align}
Next, one considers a particular bipartition of the auxiliary system $A'B'$ into $A'$ and $B'$ which satisfies the following requirements on the various PEEs
\begin{align}
	s_{AA'}(A)=s_{BB'}(B)~~,~~~ s_{AA'}(A')=s_{BB'}(B')\,,\label{balance-condition}
\end{align}
termed as the \textit{balance conditions} \cite{Wen:2021qgx,Camargo:2022mme}. Under these balance conditions, the BPE is defined as the PEE $s_{AA'}(A)$ minimized over all possible partitions that satisfy the balance requirements:
\begin{align}
	\text{BPE}(A:B)=\underset{\psi}{\text{min}}\,\,s_{AA'}(A)\Big|_{\text{balance}}\,.
\end{align}
With the above definition, the BPE may be considered as a generalization of the reflected entropy \cite{Dutta:2019gen} to general purifications. It was demonstrated in \cite{Wen:2021qgx} that the canonical purification procedure for the reflected entropy naturally satisfies the balance requirements \cref{balance-condition}.
Similar to the reflected entropy, BPE quantifies classical correlations along with quantum entanglement. It was also shown in \cite{Wen:2021qgx} that for holographic theories, the BPE exactly equals the minimal entanglement wedge cross-section(EWCS) which strongly suggests a corresponding holographic duality. This duality was further tested in a recent communication \cite{Camargo:2022mme} in the context of $3d$ flat space holography for dual BMS field theories.
Furthermore, the crossing correlation between $A$ and $B'$ and $B$ and $A'$ were also studied in \cite{Wen:2021qgx,Camargo:2022mme}. In particular, as described in \cite{Wen:2021qgx} the crossing PEEs provide the natural generalization of the so called Markov gap \cite{Hayden:2021gno} and for mixed state configurations involving adjacent intervals in holographic CFT$_2$s, the crossing PEEs turn out to be universal.

\section{BPE in Flat Holography}
\label{sec:BPE-flat}
In this section we compute the BPE for bipartite mixed states in the GCFT$_2$s dual to bulk geometries governed by TMG in asymptotically flat spacetimes. In the following we first briefly review the basic notions of the holographic duality in asymptotically flat spacetimes and the geometric picture of the entanglement structure in these flat holographic settings. Subsequent to this, we provide the computations of the BPE as well as the crossing correlations for bipartite states involving adjacent and non-adjacent intervals in GCFT$_2$s dual to Minkowski space, the global orbifold and the flat space cosmological (FSC) geometry respectively. In particular, for adjacent intervals we will also investigate the crossing PEEs as Markov gaps and find that for purely Einstein gravity in the bulk the Markov gap vanishes identically indicating signatures of a perfect Markov recovery.
\subsection{Entanglement in Galilean conformal field theory and flat holography}
\label{sec:flat-holography}
In this subsection, we will briefly recapitulate the salient features of the holographic duality between $(1+1)$-dimensional Galilean conformal field theories (GCFT$_2$) \cite{GCA,Bagchi:2009pe,Correlation,Bagchi2015EntanglementEI,Basu_2016, Log,Alishahiha:2009np,Nishida:2007pj,Bagchi:2010vw,Hosseini:2015uba,Lukierski:2005xy,PhysRevD.5.377,0264-9381-10-11-006,Martelli:2009uc,Duval:2009vt} and topologically massive gravity (TMG) \cite{Deser1982ThreeDimensionalMG,Deser1982TopologicallyMG,Jiang:2017ecm,Skenderis_2009,Hijano:2017eii,Castro_2014} in asymptotically flat spacetimes. We will also review the geometric construction of holographic entanglement in such non-relativistic field theories described in \cite{Jiang:2017ecm,Hijano:2017eii,Basu:2021awn}. The bulk action of the TMG involves a gravitational Chern-Simons (CS) term coupled with the usual Einstein-Hilbert (EH) action \cite{Jiang:2017ecm,Castro_2014} which introduces a topological contribution to the holographic entanglement entropy as discussed in \cite{Jiang:2017ecm,Basu:2021axf}. 

To begin with, we recall that the Galilean conformal algebra (GCA$_2$) describing the GCFT$_2$s may be obtained from a non-relativistic In\"on\"u-Wigner contraction of the relativistic Virasoro algebra as 
\begin{align}
	t\to t~~,~~x\to\epsilon x\,,
\end{align}
with $\epsilon\to 0$. For a holographic GCFT$_2$ located at the null infinity of the asymptotically flat spacetime, the asymptotic symmetry algebra is isomorphic to GCA$_2$ as follows
\begin{align}
	&[L_m,L_n]=(m-n)L_{m+n}+\frac{c_L}{12}n(n^2-1)\delta_{n+m,0}\,,\notag\\
	&[L_m,M_n]=(m-n)M_{m+n}+\frac{c_M}{12}n(n^2-1)\delta_{n+m,0}\,,\notag\\
	&[M_m,M_n]=0\,,
\end{align}
where $L_n\,,\,M_n$ are the generators of the GCA$_2$. For bulk dual spacetimes involving TMG, the central charges\footnote{Note that the central charges for the two sectors of the GCFT$_2$ are in general not equal. } are given by \cite{Hijano:2017eii,Bagchi:2010vw,Log,Jiang:2017ecm}
\begin{align}
	c_L=\frac{3}{\mu G_N}~~,~~c_M=\frac{3}{G_N}\,,\label{central-charges}
\end{align}
where $\mu$ denotes the coupling constant of the CS term with the EH action. For very weak coupling we have $\mu\to\infty$ and one recovers Einstein gravity with $c_L=0$ \cite{Hijano:2017eii,Bagchi:2010vw,Log,Jiang:2017ecm}, as seen from \cref{central-charges}. In this article, we consider classical solutions to TMG without cosmological constant which takes the following form in the Bondi gauge \cite{Hijano:2017eii,Log,Jiang:2017ecm}
\begin{align}
	ds^2=8G_N M\,du^2-2du\,dr+8G_N J du\,d\phi+r^2d\phi^2,\label{gen-sol-metric}
\end{align}
where $M$ is the ADM mass, $J$ is related to the angular momentum, $r$ denotes the length along the holographic direction and the coordinates $(u,\phi)$ describe the dual field theory on the asymptotic boundary.

The entanglement entropy in the above flat holographic setup was investigated in \cite{Bagchi2015EntanglementEI,Basu_2016,Jiang:2017ecm,Hijano:2017eii,Hijano:2018nhq}. Note that due to lack of Lorentz invariance, the entanglement entropy in the dual GCFT$_2$s depends on the choice of the frame. This requires considering Galilean boosted intervals and hence the scenario becomes time-dependent. A covariant geometric construction for the holographic entanglement entropy was formulated in \cite{Jiang:2017ecm,Hijano:2017eii} in the spirit of the HRT formula in usual AdS/CFT. This novel geometric picture involves two null geodesics $\gamma_i$ emanating from the endpoints $\partial_iA$ of the interval $A$ and a third extremal spacelike curve $\mathcal{E}_A$ connecting the two null lines. According to the flat space analogue of the HRT formula \cite{Jiang:2017ecm,Hijano:2017eii}, the length of the spacelike curve $\mathcal{E}_A$ computes the entanglement entropy of $A$ in units of the Planck length $\ell_P=4G_N$. Furthermore, in gravitational dual theories involving TMG, the entanglement entropy receives a topological contribution originating from the \textit{twist} in the extremal curve \cite{Jiang:2017ecm,Hijano:2017eii,Basu:2021axf} induced from the Chern-Simons part of the bulk action. Utilizing a generalization of the gravitational path integral prescription in \cite{Lewkowycz:2013nqa}, it was demonstrated in \cite{Wen:2018mev} that this geometric construction for the entanglement entropy remains valid for generic spacetimes with non-Lorentz invariant duals. Inspired by the above developments, a geometric construction for the entanglement wedge and subsequently a prescription to compute the minimal EWCS for the connected wedge configuration was put forward in \cite{Basu:2021awn}. 
Recently, the authors in \cite{Basak:2022cjs,Setare:2022uid} computed the reflected entropy in dual GCFT$_2$s and found agreements with the results of \cite{Basu:2021awn}.

In the following we will compute the BPE in such flat holographic settings for adjacent and non-adjacent intervals in the dual GCFT$_2$s and observe that our results match exactly with the minimal EWCS computed in \cite{Basu:2021awn}. We will also compare our results with the reflected entropy computed through suitable replica techniques devised in \cite{Basak:2022cjs,Setare:2022uid} and find agreement upto some constant, owing to the undetermined OPE coefficients.

\subsection{BPE for adjacent intervals}
\subsubsection{Pure Minkowski spacetime}
\begin{figure}[ht]
	\centering
	\includegraphics[scale=0.7]{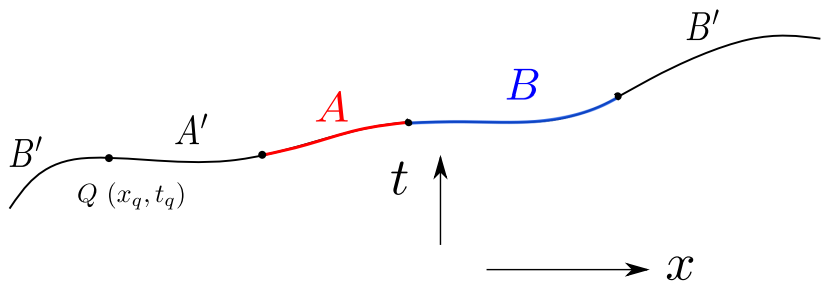}
	\caption{Schematics of the two adjacent intervals $A=[(x_1,t_1),(x_2,t_2)]$ and $B=[(x_2,t_2),(x_3,t_3)]$ in the ground state of a GCFT$_2$. The system is purified by introducing another system $A'B'$ partitioned at the point $Q:(x_q,t_q)$.}
	\label{fig:adj}
\end{figure}

We begin with the bipartite mixed state configuration described by two adjacent intervals $A\equiv[(x_1,t_1),(x_2,t_2)]$ and $B\equiv[(x_2,t_2),(x_3,t_3)]$ in the ground state of a GCFT$_{1+1}$ whose gravity dual is described by TMG in pure Minkowski spacetime\footnote{Note that we have used the so called \textit{plane representation} \cite{Bagchi:2009pe,Correlation} of the GCFT$_2$ vacuum state described by the planar coordinates $(x,t)$, which are related to the familiar cylindrical coordinates $(u,\phi)$ via a Galilean conformal transformation \cite{Log}.}. The Minkowski spacetime corresponds to the null orbifold $M=J=0$ of the solution \cref{gen-sol-metric} and has a degenerate metric given by \cite{Jiang:2017ecm,Hijano:2017eii}
\begin{align}
	ds^2=-2\,du\,dr+r^2\,d\phi^2\label{Mink-metric}\,.
\end{align}
The entanglement entropies for the intervals $A$ and $B$ are given by \cite{Jiang:2017ecm,Hijano:2017eii}
\begin{align}
	S(A)=\frac{c_L}{6}\log\left(\frac{t_{12}}{\epsilon}\right)+\frac{c_M}{6}\frac{x_{12}}{t_{12}}~~,~~S(B)=\frac{c_L}{6}\log\left(\frac{t_{23}}{\epsilon}\right)+\frac{c_M}{6}\frac{x_{23}}{t_{23}}\,,\label{EE-Mink1}
\end{align}
where $t_{ij}=t_i-t_j$ and $x_{ij}-x_i-x_j$.
Next we purify the mixed state under consideration by introducing two auxiliary intervals $A'$ and $B'$ adjacent to $A$ and $B$ as shown in \cref{fig:adj}. Furthermore, we assume that the auxiliary system $A'\cup B'$ is partitioned at a point $Q: (x_q,t_q).$ The entanglement entropies of these auxiliary intervals are given as
\begin{align}
	S(A')=\frac{c_L}{6}\log\left(\frac{t_{q1}}{\epsilon}\right)+\frac{c_M}{6}\frac{x_{q1}}{t_{q1}}~~,~~S(B')=\frac{c_L}{6}\log\left(\frac{t_{3q}}{\epsilon}\right)+\frac{c_M}{6}\frac{x_{3q}}{t_{3q}}\,.\label{EE-Mink2}
\end{align}
We will utilize the above expressions for the entanglement entropies in the balance conditions to obtain the exact position of the partition point $Q$ and subsequently compute the BPE for the mixed state configuration of two adjacent intervals. 

\subsubsection*{Einstein gravity}
We begin with pure Einstein gravity in the bulk for which the dual GCFT$_2$ has a vanishing central charge, $c_L=0$. Utilizing the PEE proposal in \cref{PEE-proposal}, the balance conditions given in \cref{balance-condition} reduce to the following relation 
\begin{align}
	\frac{x_{12}}{t_{12}}-\frac{x_{23}}{t_{23}}=\frac{x_{q1}}{t_{q1}}-\frac{x_{3q}}{t_{3q}}\label{Balance}\,.
\end{align}
We may solve the above equation for the parameter $x_q$ as follows
\begin{align}
	x_q=\frac{t_{q1} t_{3q}}{t_{13}} \left(\frac{x_{12}}{t_{12}}+\frac{x_1}{t_{1q}}+\frac{x_{23}}{t_{23}}-\frac{x_3}{t_{3q}}\right)\,.\label{Balance-soln}
\end{align}
This relation still has one undetermined parameter $t_q$. We will find that the all such solutions of the form \cref{Balance-soln} lead to the same BPE. With $Q$ undetermined, the PEE $s_{AA'}(A)$ is obtained from \cref{PEE-proposal} as
\begin{align}
	s_{AA'}(A)=\frac{1}{2}\left(S(A)+S(A\cup A')-S(A')\right)=\frac{c_M}{12}\left(\frac{x_{12}}{t_{12}}+\frac{x_{q2}}{t_{q2}}-\frac{x_{q1}}{t_{q1}}\right)\,.
\end{align}
Utilizing the solution to the balancing condition in \cref{Balance-soln} in the above expression for the PEE, we may obtain the BPE to be
\begin{align}
	\text{BPE}(A:B)=\frac{c_M}{12}\left(\frac{x_{12}}{t_{12}}+\frac{x_{23}}{t_{23}}-\frac{x_{13}}{t_{13}}\right)\,.\label{BPE-Mink1}
\end{align}
Interestingly, the BPE obtained above matches exactly with the EWCS for the two adjacent intervals $A$ and $B$ in the vacuum state of the GCFT$_2$ which is dual to Einstein gravity in Minkowski spacetime, obtained in \cite{Basu:2021awn}. This serves as a consistency check of our holographic construction for the BPE.
 
We may also compute the crossing PEE for this configuration using \cref{PEE-proposal,EE-Mink1,EE-Mink2} as follows
\begin{align}
	\mathcal{I}(A,B')=\frac{1}{2}\left(S(A'A)+S(AB)-S(A')-S(B)\right)=\frac{c_M}{12}\left(\frac{x_{q2}}{t_{q2}}+\frac{x_{13}}{t_{13}}-\frac{x_{q1}}{t_{q1}}-\frac{x_{23}}{t_{23}}\right)\,.\label{cPEE-generic}
\end{align}
It is easy to verify that under the balance condition the above crossing PEE vanishes identically\footnote{Note that similar observations were made in \cite{Camargo:2022mme} in the context of adjacent intervals in the ground state of a BMS field theory dual to flat space Einstein gravity in Minkowski spacetime. Interestingly, the authors in \cite{Camargo:2022mme} obtained a similar expression for the BPE in \cref{BPE-Mink1} which may be related to our result utilizing the BMS$_3$/GCA$_2$ correspondence \cite{Bagchi2012BMSGCART,Fareghbal:2013ifa}.} indicating a perfect Markov recovery \cite{Hayden:2021gno,Camargo:2022mme}
\begin{align}
	\text{PEE}(A:B)=\mathcal{I}(A,B')\Big|_{\text{balance}}=0\,.
\end{align}
\subsubsection*{Topologically massive gravity}
Next we turn our attention towards the bulk gravitational theory involving TMG in the null orbifold for which the dual GCFT$_2$ has a non-vanishing $c_L$. In this case, the balancing condition is modified to include the topological Chern-Simons contributions to the entropies as follows
\begin{align}
	\frac{c_L}{6}\log\left(\frac{t_{12}}{t_{23}}\right)+\frac{c_M}{6}\left(\frac{x_{12}}{t_{12}}-\frac{x_{23}}{t_{23}}\right)=\frac{c_L}{6}\log\left(\frac{t_{q1}}{t_{3q}}\right)+\frac{c_M}{6}\left(\frac{x_{q1}}{t_{q1}}-\frac{x_{3q}}{t_{3q}}\right)\,.
\end{align}
Since the central charges $c_L$ and $c_M$ are independent, this leads to the following additional constraint equation along with \cref{Balance} arising from the topological Chern-Simons contributions:
\begin{align}
	t_q=-\frac{t_1 t_2+t_3 t_2-2 t_1 t_3}{t_1-2 t_2+t_3}\,.\label{Balance-cond-2}
\end{align}
Note that this additional condition determines the partition point $Q$ uniquely unlike the Einstein gravity case where we obtained a curve instead. The PEE in this case is given by
\begin{align}
	s_{AA'}(A)=\frac{c_L}{12}\log\left(\frac{t_{12}t_{q2}}{\epsilon\, t_{q1}}\right)+\frac{c_M}{12}\left(\frac{x_{12}}{t_{12}}+\frac{x_{q2}}{t_{q2}}-\frac{x_{q1}}{t_{q1}}\right)\,.
\end{align}
Utilizing the balance conditions in \cref{Balance-cond-2,Balance-soln}, the BPE in this case becomes
\begin{align}
		\text{BPE}(A:B)=\frac{c_L}{12}\log\left[\frac{2\, t_{12}t_{23}}{\epsilon\, (t_{12}+t_{23})}\right]+\frac{c_M}{12}\left(\frac{x_{12}}{t_{12}}+\frac{x_{23}}{t_{23}}-\frac{x_{13}}{t_{13}}\right)\,,
\end{align}
which matches exactly with the EWCS for the mixed state configuration under consideration as obtained in \cite{Basu:2021awn}. We may also compare this result with the reflected entropy computed in \cite{Basak:2022cjs,Setare:2022uid} through field theoretic replica techniques and find agreement upto a constant stemming from the undetermined OPE coefficient for the CGA$_2$ three-point function.
This further substantiates the claim for a holographic duality between the EWCS and the BPE.

Furthermore, under the balance condition the crossing PEE turns out to be a non-vanishing universal constant proportional to $c_L$ in case of TMG in the bulk dual geometry as follows
\begin{align}
	\text{PEE}(A:B)=\mathcal{I}(A,B')\Big|_{\text{balance}}=\frac{c_L}{12}\log\left(\frac{t_{q2}t_{13}}{t_{q1}t_{23}}\right)=\frac{c_L}{12}\log 2\,.
\end{align}
This is reminiscent of the geometric interpretation of the Markov gap in terms of the number of boundaries of the EWCS \cite{Hayden:2021gno,Basu:2021awn} on using the flat space analogues of the Brown-Henneaux relation in \cref{central-charges}. With this result as a motivation, we now propose the following bound on the value of the crossing PEE at the balance point similar to the AdS/CFT scenario
\begin{align}
	\text{crossing PEE}(A:B)\big|_{\text{balance}}\geq \frac{\log 2}{4\mu G_N}|\partial \Sigma_{AB}|\,,\label{Bound}
\end{align}
where $|\partial \Sigma_{AB}|$ counts the non-trivial boundaries of the EWCS. Note that the above proposal already accounts for the vanishing of the crossing PEEs for purely Einstein gravity in the bulk for which we have $\mu\to\infty$. Furthermore, the above bound is saturated for the mixed state configuration of two adjacent intervals in dual GCFT$_2$s.

\subsubsection{Global Minkowski orbifold}
In this subsection, we focus on a GCFT$_2$ with a finite size defined on a cylinder of circumference $L$ which is compactified in the spatial direction described by the coordinate $\phi$. The bulk dual spacetime is described by the global Minkowski orbifold which may be obtained from the general solution to the Einstein equations in \cref{gen-sol-metric} with the following values of the ADM mass and angular momentum
\begin{align}
	M=-\left(\frac{2\pi}{L}\right)^2~~,~~J=0\,.
\end{align}
For an interval $D\equiv[(x_i,t_i),(x_j,t_j)]$ in the GCFT$_2$ with a finite size, the entanglement entropy is given by \cite{Jiang:2017ecm,Hijano:2017eii,Basu:2021axf}
\begin{align}
	S(D)=\frac{c_L}{6}\log\left[\frac{L}{\pi\epsilon}\sin\left(\frac{\pi\phi_{ij}}{L}\right)\right]+\frac{c_M}{6}\frac{\pi u_{ij}}{L}\cot\left(\frac{\pi\phi_{ij}}{L}\right)\,.\label{EE-GMO}
\end{align}
where $u_{ij}=u_i-u_j$ and $\phi_{ij}=\phi_i-\phi_j$. As earlier we consider the mixed state configuration described by two adjacent intervals $A=[(u_1,\phi_1),(u_2,\phi_2)]$ and $B=[(u_2,\phi_2),(u_3,\phi_3)]$ with $u_3>u_2>u_1$ to keep the degrees of freedom in an order. Furthermore, we purify the mixed state $\rho_{AB}$ by introducing two auxiliary intervals $A'=[(u_q,\phi_q),(u_1,\phi_1)]$ and $B'=(A\cup b\cup A')^c$ partitioned at the point $Q: (u_q,\phi_q)$. The balance conditions in \cref{balance-condition} determine the position of the partition point for the balanced minimal purification.
\subsubsection*{Einstein gravity}
First we consider the case of purely Einstein gravity in the bulk for which the dual GCFT$_2$ has $c_L=0$. With the expression for the entaglement entropy given in \cref{EE-GMO}, the balance condition in \cref{balance-condition} leads to 
\begin{align}
	\frac{u_{12}}{\tan \left(\frac{\pi \phi_{12}}{L }\right)}-\frac{u_{23}}{\tan \left(\frac{\pi \phi_{23}}{L }\right)}=\frac{u_{q1}}{\tan \left(\frac{\pi  \phi_{q1}}{L }\right)}-\frac{u_{3q}}{\tan \left(\frac{\pi  \phi_{3q}}{L }\right)}\,.\label{BC-GMO1}
\end{align}
The solution to the above constraint equation is given as
\begin{align}
	u_q= \frac{u_{12} \cot \left(\frac{\pi  \phi_{12}}{L}\right)+u_1 \cot \left(\frac{\pi \phi_{q1}}{L}\right)-u_{23} \cot \left(\frac{\pi  \phi_{23}}{L}\right)+u_3 \cot \left(\frac{\pi  \phi_{3q}}{L}\right)}{\cot \left(\frac{\pi \phi_{q1}}{L}\right)+\cot \left(\frac{\pi \phi_{3q}}{L}\right)}\,.\label{BC-sol-GMO1}
\end{align}
Note that the above solution still involves one undetermined parameter $\phi_q$ as earlier and therefore describes the equation of a line. Nevertheless, we will find that all of the points on this curve give rise to the same BPE in the case of Einstein gravity in the bulk. The PEE in this case is give by
\begin{align}
	s_{AA'}(A)=\frac{c_M}{12}\left[\frac{u_{12}}{\tan \left(\frac{\pi \phi_{12}}{L}\right)}+\frac{u_{q2}}{\tan \left(\frac{\pi \phi_{q2}}{L}\right)}-\frac{u_{q1}}{\tan \left(\frac{\pi \phi_{q1}}{L}\right)}\right]\,.\label{PEE-GMO1}
\end{align}
Now substituting the solution in \cref{BC-sol-GMO1} to the balance condition into \cref{PEE-GMO1}, we uniquely determine the BPE for this configuration as
\begin{align}
	\text{BPE}(A:B)=\frac{c_M}{12}\left[\frac{u_{12}}{\tan \left(\frac{\pi \phi_{12}}{L}\right)}+\frac{u_{23}}{\tan \left(\frac{\pi \phi_{23}}{L}\right)}-\frac{u_{13}}{\tan \left(\frac{\pi \phi_{13}}{L}\right)}\right]\,.
\end{align}
Once again the BPE obtained above matches exactly with the minimal EWCS for the present configuration of two adjacent intervals in a finite sized GCFT$_2$ dual to Einstein gravity in the global Minkowski orbifold.

The crossing PEE is obtained from \cref{cPEE-generic} as
\begin{align}
		\mathcal{I}(A,B')=\frac{c_M}{12}\left[\frac{u_{q2}}{\tan \left(\frac{\pi \phi_{q2}}{L}\right)}+\frac{u_{13}}{\tan \left(\frac{\pi \phi_{13}}{L}\right)}-\frac{u_{q1}}{\tan \left(\frac{\pi L}{\beta }\right)}-\frac{u_{23}}{\tan \left(\frac{\pi  \phi_{23}}{L}\right)}\right]\,.
\end{align}
Once again it is easy to verify that at the balance point given by \cref{BC-sol-GMO1}, the above crossing PEE vanishes identically. This provides further evidence to the fact that there is a perfect Markov recovery in GCFT$_2$s dual to Einstein gravity.
\subsubsection*{Topologically massive gravity}
In the case of TMG in global Minkowski orbifolds, we have both central charges of the dual non-relativistic field theory to be non-vanishing as seen from \cref{central-charges}. Therefore, utilizing \cref{balance-condition,EE-GMO} the balance condition in \cref{balance-condition} becomes
\begin{align}
		\frac{c_L}{6}\log\left[\frac{\sin\left(\frac{\pi \phi_{12}}{L}\right)}{\sin\left(\frac{\pi \phi_{23}}{L }\right)}\right]+&\frac{c_M}{6}\left[\frac{u_{12}}{\tan \left(\frac{\pi \phi_{12}}{L}\right)}-\frac{u_{23}}{\tan \left(\frac{\pi \phi_{23}}{L }\right)}\right]\notag\\=&\frac{c_L}{6}\log\left[\frac{\sin\left(\frac{\pi \phi_{q1}}{L}\right)}{\sin\left(\frac{\pi \phi_{3q}}{L }\right)}\right]+\frac{c_M}{6}\left[\frac{u_{q1}}{\tan \left(\frac{\pi \phi_{q1}}{L}\right)}-\frac{u_{3q}}{\tan \left(\frac{\pi \phi_{3q}}{L}\right)}\right]\,.
\end{align}
The above balance condition implies one more constraint equation along with \cref{BC-GMO1} obtained in the context of Pure Einstein gravity. This is given as follows
\begin{align}
	\frac{\sin\left(\frac{\pi \phi_{12}}{L }\right)}{\sin\left(\frac{\pi \phi_{23}}{L }\right)}=-\frac{\sin\left(\frac{\pi \phi_{q1}}{L }\right)}{\sinh\left(\frac{\pi \phi_{3q}}{L }\right)}\,,\label{BC-FSC2}
\end{align}
solving which we obtain the remaining parameter $\phi_q$ as follows
\begin{align}
	\phi_q= 2\,\arccos\left[\frac{\sqrt{2} \left(2 \cos \left(\frac{\pi\phi_1}{L}\right) \sin \left(\frac{\pi\phi_2}{L}\right) \cos \left(\frac{\pi\phi_3}{L}\right)-\cos \left(\frac{\pi\phi_2}{L}\right) \sin \left(\frac{\pi(\phi_1+\phi_3)}{L}\right)\right)}{\sqrt{3-2 \cos \left(\frac{\phi_{12}}{L}\right)+\cos \left(\frac{\phi_{13}}{L}\right)-2 \cos \left(\frac{\phi_{23}}{L}\right)}}\right]\,.\label{BC-sol-GMO2}
\end{align}
Note that the solutions in \cref{BC-sol-GMO1,BC-sol-GMO2} uniquely determine the partition point $Q$. Now substituting the solutions to the balance conditions given in \cref{BC-sol-GMO1,BC-sol-GMO2} in the expression for the partial entanglement entropy $s_{AA'}(A)$, we obtain the following result for the BPE for the present mixed state configuration
\begin{align}
	\text{BPE}(A:B)=\frac{c_L}{12}\log \left[\frac{2 L}{\pi\epsilon}\,\frac{\sin \left(\frac{\pi \phi_{12}}{L }\right)\sin \left(\frac{\pi  \phi_{23}}{L}\right)}{\sin \left(\frac{\pi  \phi_{13}}{L}\right)}\right]+\frac{c_M}{12}&\Bigg[\frac{u_{12}}{\tan \left(\frac{\pi \phi_{12}}{L}\right)}\notag\\
	&+\frac{u_{23}}{\tan \left(\frac{\pi \phi_{23}}{L}\right)}-\frac{u_{13}}{\tan \left(\frac{\pi \phi_{13}}{L}\right)}\Bigg]\,.
\end{align}
This expression for the BPE for the mixed state configuration of two adjacent intervals in the GCFT$_2$ dual to TMG in global Minkowski orbifold exactly matches with the corresponding minimal EWCS obtained through the geometric construction in \cite{Basu:2021awn}. The above BPE is also connotative of the  field theory replica technique result for the reflected entropy for the present configuration \cite{Basak:2022cjs,Setare:2022uid}.
Finally, we compute the crossing PEE at the balance point which again turns out to be an universal constant proportional to $c_L$ as follows
\begin{align}
	\text{PEE}(A:B)=\mathcal{I}(A,B')\Big|_{\text{balance}}=\frac{c_L}{12}\log\left[
	\frac{\sin \left(\frac{\pi \phi_{q2}}{L}\right)\sin \left(\frac{\pi  \phi_{13}}{L}\right)}{\sin \left(\frac{\pi  \phi_{q1}}{L}\right) \sin\left(\frac{\pi  \phi_{23}}{L }\right)}\right]=\frac{c_L}{12}\log 2\,.
\end{align}
Utilizing the value of the central charge $c_L$ in \cref{central-charges}, the above expression for the crossing PEE saturates the proposed bound in \cref{Bound} and corresponds to the geometric interpretation of the Markov gap in terms of the number of boundaries of the EWCS similar to the usual AdS/CFT scenario \cite{Hayden:2021gno,Wen:2021qgx}.
\subsubsection{Flat space cosmology}
Next we consider the case of a thermal GCFT$_2$ defined on a cylinder of circumference $\beta$ which is compactified in the timelike direction. The dual bulk spacetime is described by TMG in the non-rotating ($J=0$) flat space cosmological (FSC) geometry with the metric
\begin{align}
	ds^2=M\,du^2-2\,du\,dr+r^2d\phi^2\,.
\end{align}
The ADM mass of the spacetime $M$ is related to the temperature in the dual field theory as
\begin{align}
	M=\left(\frac{2\pi}{\beta}\right)^2\,.
\end{align}
The entanglement entropy for an interval $D\equiv[(\rho_i,\phi_i),(\rho_j,\phi_j)]$ in such a thermal GCFT$_2$ located at the asymptotic boundary of the FSC spacetime is given by
\begin{align}
	S(D)=\frac{c_L}{6}\log\left[\frac{\beta}{\pi\epsilon}\sinh\left(\frac{\pi\phi_{ij}}{\beta}\right)\right]+\frac{c_M}{6}\frac{\pi u_{ij}}{\beta}\coth\left(\frac{\pi\phi_{ij}}{\beta}\right)\,.\label{EE-FSC}
\end{align}
where $u_{ij}=u_i-u_j$ and $\phi_{ij}=\phi_i-\phi_j$. Once again, we consider two adjacent intervals  $A=[(u_1,\phi_1),(u_2,\phi_2)]$ and $B=[(u_2,\phi_2),(u_3,\phi_3)]$ with $u_3>u_2>u_1$. The mixed state $\rho_{AB}$ is purified by an auxiliary system $A'B'$ partitioned at the point $Q:(u_q,\phi_q)$.
\subsubsection*{Einstein gravity}
We begin with the case of pure Einstein gravity in the bulk geometry for which the dual field theory has $c_L=0$. Using \cref{EE-FSC} in the balance condition given in \cref{balance-condition}, we obtain
\begin{align}
	\frac{u_{12}}{\tanh \left(\frac{\pi \phi_{12}}{\beta }\right)}-\frac{u_{23}}{\tanh \left(\frac{\pi \phi_{23}}{\beta }\right)}=\frac{u_{q1}}{\tanh \left(\frac{\pi  \phi_{q1}}{\beta }\right)}-\frac{u_{3q}}{\tanh \left(\frac{\pi  \phi_{3q}}{\beta }\right)}\,.\label{BC-FSC1}
\end{align}
The above constraint equation has two undetermined parameters $u_q$ and $\phi_q$ which specify the partition point. We can now solve this equation for the parameter $u_q$ as follows
\begin{align}
u_q=  \frac{\sinh \left(\frac{\pi \phi_{3q}}{\beta }\right) \sinh \left(\frac{\pi  \phi_{q2}}{\beta }\right)}{\sinh\left(\frac{\pi \phi_{12}}{\beta }\right) \sinh\left(\frac{\pi \phi_{31}}{\beta }\right)}\,u_1+\frac{\sinh \left(\frac{\pi \phi_{3q}}{\beta }\right) \sinh \left(\frac{\pi  \phi_{q1}}{\beta }\right)}{\sinh\left(\frac{\pi \phi_{12}}{\beta }\right) \sinh\left(\frac{\pi \phi_{23}}{\beta }\right)}\,u_2+\frac{\sinh \left(\frac{\pi \phi_{2q}}{\beta }\right) \sinh \left(\frac{\pi  \phi_{q1}}{\beta }\right)}{\sinh\left(\frac{\pi \phi_{23}}{\beta }\right) \sinh\left(\frac{\pi \phi_{31}}{\beta }\right)}\,u_3\,. \label{BC-sol-FSC1}
\end{align}
Note that the above solution still depends on the undetermined parameter $\phi_q$ and therefore denotes the equation of a curve. Remarkably, all the points on this curve lead to the same value of the BPE. Utilizing the expression for the entropy given in \cref{EE-FSC}  in the PEE proposal \cref{PEE-proposal}, we obtain the PEE $s_{AA'}(A)$ from \cref{cPEE-generic} as
\begin{align}
	s_{AA'}(A)=\frac{c_M}{12}\left[\frac{u_{12}}{\tanh \left(\frac{\pi \phi_{12}}{\beta }\right)}+\frac{u_{q2}}{\tanh \left(\frac{\pi \phi_{q2}}{\beta }\right)}-\frac{u_{q1}}{\tanh \left(\frac{\pi \phi_{q1}}{\beta }\right)}\right]\,.
\end{align}
Substituting the solution in \cref{BC-sol-FSC1} to the balance condition in the above expression, the BPE for the mixed state configuration of two adjacent intervals in the thermal GCFT$_2$ is obtained to be
\begin{align}
	\text{BPE}(A:B)=\frac{c_M}{12}\left[\frac{u_{12}}{\tanh \left(\frac{\pi \phi_{12}}{\beta }\right)}+\frac{u_{23}}{\tanh \left(\frac{\pi \phi_{23}}{\beta }\right)}-\frac{u_{13}}{\tanh \left(\frac{\pi \phi_{13}}{\beta }\right)}\right]\,.\label{BPE-FSC}
\end{align}
Interestingly, the BPE obtained above once again matches exactly with the minimal EWCS for the present configuration obtained in \cite{Basu:2021awn} through the novel geometric construction. Furthermore, \cref{BPE-FSC} agrees with the reflected entropy for the two adjacent intervals in the thermal GCFT$_2$ obtained in \cite{Basak:2022cjs,Setare:2022uid}. This provides further support to the holographic duality between all these quantities in the flat holographic setting.

We may also compute the crossing PEE from \cref{cPEE-generic,EE-FSC} as follows
\begin{align}
	\mathcal{I}(A,B')=\frac{c_M}{12}\left[\frac{u_{q2}}{\tanh \left(\frac{\pi \phi_{q2}}{\beta }\right)}+\frac{u_{13}}{\tanh \left(\frac{\pi \phi_{13}}{\beta }\right)}-\frac{u_{q1}}{\tanh \left(\frac{\pi  \phi_{q1}}{\beta }\right)}-\frac{u_{23}}{\tanh \left(\frac{\pi  \phi_{23}}{\beta }\right)}\right]\,.
\end{align}
It is easy to verify that under the balance condition the crossing PEE vanishes identically. Therefore, once again we establish the signatures of a perfect Markov recovery in the case of GCFT$_2$s with pure Einstein gravity duals.
\subsubsection*{Topologically massive gravity}
Next we move on to the computation of the BPE for the two adjacent intervals $A$ and $B$ in the thermal GCFT$_2$ with both the central charges being non-zero. The gravity dual is described by TMG in the FSC geometry and the entanglement entropy for a generic interval in the dual field theory picks up a topological Chern-Simons contribution as in \cref{EE-FSC}. For this case the balance condition in \cref{balance-condition} becomes
\begin{align}
	\frac{c_L}{6}\log\left[\frac{\sinh\left(\frac{\pi \phi_{12}}{\beta }\right)}{\sinh\left(\frac{\pi \phi_{23}}{\beta }\right)}\right]+&\frac{c_M}{6}\left[\frac{u_{12}}{\tanh \left(\frac{\pi \phi_{12}}{\beta }\right)}-\frac{u_{23}}{\tanh \left(\frac{\pi \phi_{23}}{\beta }\right)}\right]\notag\\=&\frac{c_L}{6}\log\left[\frac{\sinh\left(\frac{\pi \phi_{q1}}{\beta }\right)}{\sinh\left(\frac{\pi \phi_{3q}}{\beta }\right)}\right]+\frac{c_M}{6}\left[\frac{u_{q1}}{\tanh \left(\frac{\pi \phi_{q1}}{\beta }\right)}-\frac{u_{3q}}{\tanh \left(\frac{\pi \phi_{3q}}{\beta }\right)}\right]\,,
\end{align}
which imposes the following additional constraint equation along with \cref{BC-FSC1}
\begin{align}
	\frac{\sinh\left(\frac{\pi \phi_{12}}{\beta }\right)}{\sinh\left(\frac{\pi \phi_{23}}{\beta }\right)}=-\frac{\sinh\left(\frac{\pi \phi_{q1}}{\beta }\right)}{\sinh\left(\frac{\pi \phi_{3q}}{\beta }\right)}\,.\label{BC-FSC2}
\end{align}
The solution to \cref{BC-FSC2} leads to the value of the undetermined parameter $\phi_q$ as follows
\begin{align}
	\phi_q=\log \left(\frac{e^{\frac{2 \pi  (\phi_1+\phi_2)}{\beta }}-2 e^{\frac{2 \pi  (\phi_1+\phi_3)}{\beta }}+e^{\frac{2 \pi  (\phi_2+\phi_3)}{\beta }}}{2 e^{\frac{2 \pi \phi_2}{\beta }}-e^{\frac{2 \pi \phi_1}{\beta }}-e^{\frac{2 \pi \phi_3}{\beta }}}\right)\,.\label{BC-sol-FSC2}
\end{align}
The solutions in \cref{BC-sol-FSC1,BC-sol-FSC2} completely determine the partition point $Q$ which corresponds to the minimal balanced purification. Substituting the solutions in \cref{BC-sol-FSC1,BC-sol-FSC2} to the balance conditions into the corresponding expression for the PEE $s_{AA'}(A)$, the BPE is obtained as follows
\begin{align}
	\text{BPE}(A:B)=\frac{c_L}{12}\log \left[\frac{2 \,\beta}{\pi\epsilon}\frac{\sinh \left(\frac{\pi \phi_{12}}{\beta }\right)\sinh \left(\frac{\pi  \phi_{23}}{\beta }\right)}{\sinh \left(\frac{\pi  \phi_{13}}{\beta }\right)}\right]+\frac{c_M}{12}&\Bigg[\frac{u_{12}}{\tanh \left(\frac{\pi \phi_{12}}{\beta }\right)}\notag\\
	&+\frac{u_{23}}{\tanh \left(\frac{\pi \phi_{23}}{\beta }\right)}-\frac{u_{13}}{\tanh \left(\frac{\pi \phi_{13}}{\beta }\right)}\Bigg]\,.
\end{align}
The above expression matches exactly with the minimal EWCS for the two adjacent intervals in the thermal GCFT$_2$ \cite{Basu:2021awn}. Furthermore, the field theoretic replica technique calculations for the reflected entropy in \cite{Basak:2022cjs,Setare:2022uid} for the present mixed state configuration agree with our result for the BPE upto a constant providing yet another consistency check.
In this case, the crossing PEE is again a constant proportional to $c_L$ under the balance condition given by
\begin{align}
	\text{PEE}(A:B)=\frac{c_L}{12}\log\left[
	\frac{\sinh \left(\frac{\pi \phi_{q2}}{\beta }\right)\sinh \left(\frac{\pi  \phi_{13}}{\beta }\right)}{\sinh \left(\frac{\pi  \phi_{q1}}{\beta }\right) \sinh \left(\frac{\pi  \phi_{23}}{\beta }\right)}\right]=\frac{c_L}{12}\log 2\,.
\end{align}
Upon using the analogue of the Brown-Henneaux relation in \cref{central-charges}, the crossing PEE obtained above is seen to saturate the bound in \cref{Bound} and hence is consistent with the geometric interpretation of the Markov gap in terms of the number of non-trivial boundaries of the EWCS.

\subsection{BPE for non-adjacent intervals}
In this subsection, we turn our attention towards the computation of the BPE for various bipartite mixed states involving two disjoint intervals $A$ and $B$ in a GCFT$_2$ dual to TMG in asymptotically flat spacetimes. As in the adjacent cases described earlier, we consider an auxiliary system $A'B'$ to purify the mixed state $\rho_{AB}$ and subsequently require a bipartition of $A'B'$ into $A'$ and $B'$. As described in \cite{Wen:2021qgx,Camargo:2022mme}, for disjoint intervals the complement subsystem $A'B'$ becomes disconnected and one needs to partition it into $A'=A'_1\cup A'_2$ and $B'=B'_1\cup B'_2$ as shown in \cref{fig:disj}. Following the prescription in \cite{Wen:2021qgx}, the partition which leads to the minimal balanced purification turns out to be the one in which $B'$ is as far from $A$ and $A'$ is as far from $B$. This kind of partitioning of the purifier system $A'B'$ sets the disconnected pieces of $A'$ and $B'$ in natural pairs $(A'_1,B'_1)$ and $(A'_2,B'_2)$.  In this case, the generalized balance conditions are given by \cite{Wen:2021qgx,Camargo:2022mme}
\begin{align}
	s_{AA'}(A)=s_{BB'}(B)~~,~~s_{AA'}(A'_1)=s_{BB'}(B'_1)~~,~~s_{AA'}(A'_2)=s_{BB'}(B'_2)\,,
\end{align}
which reduces to the following two conditions upon utilizing the PEE proposal in \cref{PEE-proposal},
\begin{align}
	S(A'_1)-S(B'_1)=S(AA'_2)-S(BB'_2)~~,~~S(A'_2)-S(B'_2)=S(AA'_1)-S(BB'_1)\,.\label{BC-disj-gen}
\end{align}
\subsubsection{Pure Minkowski spacetime}
\begin{figure}[ht]
	\centering
	\includegraphics[scale=0.7]{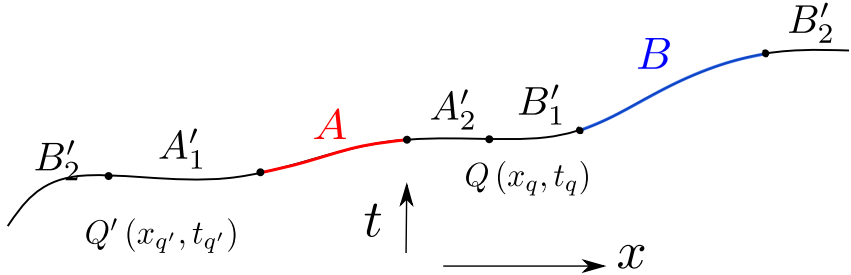}
	\caption{Schematics of the two disjoint intervals $A=[(x_1,t_1),(x_2,t_2)]$ and $B=[(x_3,t_3),(x_4,t_4)]$ in the ground state of a GCFT$_2$. The system is purified by introducing another system $A'B'$ partitioned at the points $Q:(x_q,t_q)$ and $Q':(x_{q'},t_{q'})$.}
	\label{fig:disj}
\end{figure}

We begin with the mixed state configuration of two disjoint intervals $A=[(x_1,t_1),(x_2,t_2)]$ and $B=[(x_3,t_3),(x_4,t_4)]$ in the ground state of a GCFT$_2$ which is dual to the Minkowski null orbifold with the metric given in \cref{Mink-metric}. We will further assume that $x_4>x_3>x_2>x_1$ in order to keep a sense of ordering of the subsystems. Next we partition the auxiliary purifying system $A'B'$ into $A'_1=[(x_2,t_2),(x_q,t_q)]$, $A'_2=[(x_{q'},t_{q'}),(x_1,t_1)]$, $B'_1=[(x_q,t_q),(x_3,t_3)]$ and $B'_2=(A\cup A'\cup B\cup B_1)^c$ as shown in \cref{fig:disj}. The partition points are denoted as $Q:(x_q,t_q)$ and $Q':(x_{q'},t_{q'})$. The balance condtions given in \cref{BC-disj-gen} determine the optimal positions of $Q$ and $Q'$ for the balanced purification. 

\subsubsection*{Einstein gravity}
For the case of pure Einstein gravity in the bulk, we have $c_L=0$ and the balance conditions in \cref{BC-disj-gen} imply the following constraint equations
\begin{align}
	\frac{x_{2q}}{t_{2q}}-\frac{x_{q3}}{t_{q3}}=\frac{x_{q'2}}{t_{q'2}}-\frac{x_{q'3}}{t_{q'3}}~~,~~\frac{x_{q'1}}{t_{q'1}}-\frac{x_{q'4}}{t_{q'4}}=\frac{x_{q1}}{t_{q1}}-\frac{x_{q4}}{t_{q4}}\,,\label{BC-MO-disj1}
\end{align}
where we have utilized the expression for the entanglement entropies given in \cref{EE-Mink1}. Note that we have only two constraint equations from the balance condition which are not enough to solve for the partition points $Q$ and $Q'$ uniquely. For brevity of calculations, we may set
\begin{align}
	(x_1,t_1)=(0,0)\,,\label{brevity}
\end{align}
utilizing the translational symmetry along the $(x,t)$ coordinates which describe the dual GCFT$_2$. Note that the cross-ratios corresponding to the disjoint intervals $A$ and $B$ now become
\begin{align}
	T=\frac{t_2\,(t_3-t_4)}{t_3\,(t_2-t_4)}~~,~~X=T\left(\frac{x_2}{t_2}+\frac{x_3-x_4}{t_3-t_4}-\frac{x_3}{t_3}-\frac{x_2-x_4}{t_2-t_4}\right)\,.\label{cross-ratio-PM1}
\end{align}
Solving these balance conditions in \cref{BC-MO-disj1} for the parameters $x_q$ and $x_{q'}$, we obtain
\begin{align}
	&x_q=\frac{t_4 t_q t_{4q} (t_{q3} t_{3q'} x_2 + t_{2q} t_{2q'} x_3) +  t_q t_{q'} t_{32} t_{q2} t_{q3}  x_4}{t_4 t_{23}  \left(t_2 t_3 (t_4 - t_q - t_{q'}) +  t_q t_{q'} (t_2+t_3 - t_4) \right)}\,,\notag\\
	&x_{q'}=\frac{t_4 t_{q'} t_{4q'} (t_{q3} t_{3q'} x_2 + t_{2q} t_{2q'} x_3) +  t_q t_{q'} t_{32} t_{q2} t_{q3}  x_4}{t_4 t_{23}  \left(t_2 t_3 (t_4 - t_q - t_{q'}) +  t_q t_{q'} (t_2+t_3 - t_4) \right)}\,.
	\label{BC-sol-PM-disj1}
\end{align}
 We still have two undetermined parameters $t_q$ and $t_{q'}$ which specify the two partition points $Q$ and $Q'$. The partial entanglement entropy $s_{AA'}(A)$ for the mixed state configuration of two disjoint intervals $A$ and $B$ is given by
\begin{align}
	s_{AA'}(A)&=\frac{1}{2}\left(S(AA'_1)+S(AA'_2)-S(A'_1)-S(A'_2)\right)\notag\\
	&=\frac{c_M}{12}\,\left(\frac{x_{1q}}{t_{1q}}+\frac{x_{q'2}}{t_{q'2}}-\frac{x_{2q}}{t_{2q}}-\frac{x_{q'1}}{t_{q'1}}\right)\,.\label{PEE-disj-PM}
\end{align}
Substituting the solution \cref{BC-sol-PM-disj1} to the balance conditions in the above expression for the PEE, the BPE for the present mixed state configuration reduces to
\begin{align}
BPE(A:B)=\frac{c_M}{12}\,\frac{t_{qq'}\,\big(t_3\,(t_4\,t_{34}\,x_2+t_2\,t_{23}\,x_4)-t_2\,t_4\,t_{24}\,x_3\big)}{t_4\,t_{23}\,\big[t_2\,t_3\,t_{4q}+\left((t_2+t_3-t_4)\,t_q-t_2\,t_3\right)\,t_{q'}\big]}\,.
\end{align}
The above expression still depends on both $t_q$ and $t_{q'}$. Therefore, to obtain the correct BPE we will need additional constraints along with the balance conditions in \cref{BC-MO-disj1}. To proceed, we compute the difference between the BPE obtained above and the minimal EWCS in \cite{Basu:2021awn} and find that the difference vanishes as long as the following additional conditions are satisfied
\begin{align}
	&t_q=\frac{t_2\,t_4}{(1-\sqrt{T})\,t_2+\sqrt{T}\,t_4}\,,\notag\\
	&t_{q'}=\frac{t_2\,t_4}{(1+\sqrt{T})\,t_2-\sqrt{T}\,t_4}\,.\label{auxiliary-PM}
\end{align}
Note that similar conditions\footnote{We noticed a typo in these conditions reported in \cite{Camargo:2022mme} where the square roots from the cross-ratio $T$ were missed.} were found in \cite{Camargo:2022mme} in the context of asymptotically flat spacetimes dual to BMS field theories. These auxiliary conditions lack physical background and therefore are seemingly arbitrary. In the following subsection, we will find that these auxiliary conditions naturally emerge from the topological contributions to the partial entanglement entropies when we consider the full topologically massive gravity theory dual to the GCFT$_2$ at the asymptotic boundary with both central charges non-vanishing. 

\subsubsection*{Topologically massive gravity}
For topologically massive gravity is pure Minkowski spacetime, the entanglement entropy for a generic interval on the asymptotic boundary is given in \cref{EE-GMO}. Therefore, the balance conditions in \cref{BC-disj-gen} lead to the following relations
\begin{align}
	&\frac{c_L}{6}\log\left(\frac{t_{2q}}{t_{q3}}\right)+\frac{c_M}{6}\left(\frac{x_{2q}}{t_{2q}}-\frac{x_{q3}}{t_{q3}}\right)=\frac{c_L}{6}\log\left(\frac{t_{q'2}}{t_{q'3}}\right)+\frac{c_M}{6}\left(\frac{x_{q'2}}{t_{q'2}}-\frac{x_{q'3}}{t_{q'3}}\right)\,,\notag\\
	&\frac{c_L}{6}\log\left(\frac{t_{q'1}}{t_{q'4}}\right)+\frac{c_M}{6}\left(\frac{x_{q'1}}{t_{q'1}}-\frac{x_{q'4}}{t_{q'4}}\right)=\frac{c_L}{6}\log\left(\frac{t_{q1}}{t_{q4}}\right)+\frac{c_M}{6}\left(\frac{x_{q1}}{t_{q1}}-\frac{x_{q4}}{t_{q4}}\right)\,.\label{BC-PM-disj2}
\end{align}
In this case, along with the balance constraints in \cref{BC-MO-disj1} we have the following two additional conditions
\begin{align}
	\frac{t_{2q}}{t_{q3}}=\frac{t_{q'2}}{t_{q'3}}~~,~~\frac{t_{q'1}}{t_{q'4}}=-\frac{t_{q1}}{t_{q4}}\,.
\end{align}
With the simplification in \cref{brevity} we may solve these balance conditions to obtain the following expressions for the previously undetermined parameters
\begin{align}
	&t_q=\frac{t_2 t_3 - \sqrt{t_2 t_3 (t_2 - t_4) (t_3 - t_4)}}{t_2 + t_3 - t_4}\,,\notag\\
	&t_{q'}=\frac{t_2 t_3 + \sqrt{t_2 t_3 (t_2 - t_4) (t_3 - t_4)}}{t_2 + t_3 - t_4}\,.\label{BC-sol-disj2-Mink}
\end{align}
It is straightforward to verify that these relations reduce to \cref{auxiliary-PM} upon utilizing the cross-ratio $T$ given in \cref{cross-ratio-PM1}. These solutions together with \cref{BC-sol-PM-disj1} determine the partition points $Q$ and $Q'$ completely, unlike the case with Einstein gravity in the bulk.
Substituting the solutions in \cref{BC-sol-PM-disj1,BC-sol-disj2-Mink} into the expression for the PEE $s_{AA'}(A)$ given in \cref{PEE-disj-PM} we may now obtain the BPE for the present configuration of two disjoint intervals in the ground state of the GCFT$_2$ as follows
\begin{align}
	BPE(A:B)=\frac{c_L}{12}&\log\left[\frac{-2 t_2 t_3 + 2 \sqrt{t_2 t_3 (t_2 - t_4) (t_3 - t_4)} + (t_2 + t_3) t_4}{(t_2 - t_3) t_4}\right]\notag\\
	&+\frac{c_M}{12}\,\frac{(t_3 (t_3 - t_4) t_4 x_2 + t_2 t_4 (-t_2 + t_4) x_3 + 
		t_2 (t_2 - t_3) t_3 x_4)}{(t_2 - t_3)t_4 \sqrt{t_2 t_3 (t_2 - t_4) (t_3 - t_4)} }\,.
\end{align}
where we have included the Chern-Simons contribution to the PEE as well. Now using the values of the cross-ratios given in \cref{cross-ratio-PM1}, we may rewrite the above expression in the comprehensive form
\begin{align}
	\text{BPE}(A:B)=\frac{c_L}{12}\log\left(\frac{1+\sqrt{T}}{1-\sqrt{T}}\right)+\frac{c_M}{12}\left|\frac{X}{\sqrt{T}(1-T)}\right|\,.
\end{align}
The above BPE is exactly equal to the value of the minimal EWCS for the two disjoint intervals under consideration \cite{Basu:2021awn}. Furthermore, the reflected entropy obtained in \cite{Basak:2022cjs,Setare:2022uid} matches with our result upto a constant, providing a consistency check with field theoretic methods. Therefore, incorporating the full gravitational theory involving TMG in asymptotically flat spacetimes we are able to determine the partition points uniquely. Consequently, the correct expression for the balance partial entanglement is obtained, which is equivalent to the minimal EWCS. Furthermore, we may reproduce the BPE for pure Einstein gravity in the bulk simply by taking the $c_L=0$ (or $\mu\to\infty$) limit of the TMG answer. Hence in the following, we will investigate the gravitational theories including the topological part of the action and recover the answer for pure Einstein gravity as a particular limit.

\subsubsection{Global Minkowski orbifold}
Next, we focus on the computation of the holographic BPE corresponding to two disjoint intervals  $A=[(u_1,\phi_1),(u_2,\phi_2)]$ and $B=[(u_3,\phi_3),(u_4,\phi_4)]$ in a finite sized GCFT$_2$ whose gravity dual is described by TMG in the global Minkowski orbifold. We divide the purifier system $A'B'$ into pairs by introducing partitions at two points $Q:(u_q,\phi_q)$ and $Q':(u_{q'},\phi_{q'})$, similar to the situation in the Minkowski null orbifold shown in \cref{fig:disj}.
As discussed above, it is of advantage to consider the gravitational theory including the topological term in order to have all the necessary constraint equations to solve for the partition points. The balance conditions in \cref{BC-disj-gen} lead to four different relations between the entanglement entropies of various subsystems involved. From the topological contributions to the entropies proportional to $c_L$, we obtain
\begin{align}
		\frac{\sin\left(\frac{\pi\phi_{2q}}{L}\right)}{\sin\left(\frac{\pi\phi_{q3}}{L}\right)}=\frac{\sin\left(\frac{\pi\phi_{q'2}}{L}\right)}{\sin\left(\frac{\pi\phi_{q'3}}{L}\right)}~~~,~~~\frac{\sin\left(\frac{\pi\phi_{q'}}{L}\right)}{\sin\left(\frac{\pi\phi_{q'4}}{L}\right)}=-\frac{\sin\left(\frac{\pi\phi_{q}}{L}\right)}{\sin\left(\frac{\pi\phi_{q4}}{L}\right)}\label{top-BC-GMO}\,.
\end{align}
For brevity of calculations, we utilize the conformal symmetry of the GCFT$_2$ to put one of the endpoints of the interval $A$ at the origin, namely $(u_1,\phi_1)=(0,0)$.
Now solving the above constraint equations we may determine the parameters $\phi_q$ and $\phi_{q'}$ uniquely as follows
\begin{align}
	&\phi_q=\frac{L}{\pi}\, \cot^{-1}\left[\cot\left(\frac{\pi\phi_{q'}}{L}\right)-2\,\cot\left(\frac{\pi\phi_4}{L}\right)\right]\,,\notag\\
	&\phi_{q'}=\frac{L}{\pi}\, \cot^{-1}\left[\cot\left(\frac{\pi\phi_4}{L}\right)-\csc\left(\frac{\pi\phi_4}{L}\right)\sqrt{\frac{\sin\left(\frac{\pi\phi_{24}}{L}\right)\sin\left(\frac{\pi\phi_{34}}{L}\right)}{\sin\left(\frac{\pi\phi_{2}}{L}\right)\sin\left(\frac{\pi\phi_{3}}{L}\right)}}\right]\,.\label{BC-sol-disj-GMO1}
\end{align}
Note that the above solutions do not determine the partition points completely. The topological Chern-Simons contribution to the PEE $s_{AA'}(A)$ may be written using \cref{PEE-disj-PM} as
\begin{align}
	s^{\text{CS}}_{AA'}(A)=\frac{c_L}{12}\log\left[\frac{\sin\left(\frac{\pi\phi_{1q}}{L}\right)\sin\left(\frac{\pi\phi_{q'2}}{L}\right)}{\sin\left(\frac{\pi\phi_{2q}}{L}\right)\sin\left(\frac{\pi\phi_{q'1}}{L}\right)}\right]\,.
\end{align}
Substituting the solutions \cref{BC-sol-disj-GMO1} into the above PEE, we now obtain the Chern-Simons contribution to the BPE corresponding to the disjoint intervals $A$ and $B$ in the finite sized GCFT$_2$ as follows
\begin{align}
	\text{BPE}^{\text{CS}}(A:B)=\frac{c_L}{12}\log\left[\frac{\cot\left(\frac{\pi\phi_4}{L}\right)-\cot\left(\frac{\pi\phi_2}{L}\right)-\csc\left(\frac{\pi\phi_4}{L}\right)\sqrt{\frac{\sin\left(\frac{\pi\phi_{24}}{L}\right)\sin\left(\frac{\pi\phi_{34}}{L}\right)}{\sin\left(\frac{\pi\phi_{2}}{L}\right)\sin\left(\frac{\pi\phi_{3}}{L}\right)}}}{\cot\left(\frac{\pi\phi_2}{L}\right)-\cot\left(\frac{\pi\phi_4}{L}\right)-\csc\left(\frac{\pi\phi_4}{L}\right)\sqrt{\frac{\sin\left(\frac{\pi\phi_{24}}{L}\right)\sin\left(\frac{\pi\phi_{34}}{L}\right)}{\sin\left(\frac{\pi\phi_{2}}{L}\right)\sin\left(\frac{\pi\phi_{3}}{L}\right)}}}\right]\,.
\end{align}
We may rewrite the above expression comprehensively as
\begin{align}
	\text{BPE}^{\text{CS}}(A:B)=\frac{c_L}{12}\log\left(\frac{1+\sqrt{\tilde{T}}}{1-\sqrt{\tilde{T}}}\right)\,,
\end{align}
where $\tilde{T}$ is one of the cross-ratios for the two disjoint intervals $A=[(0,0),(u_2,\phi_2)]$ and $B=[(u_3,\phi_3),(u_4,\phi_4)]$ in the dual GCFT$_2$ defined on a cylinder compactified in the spatial direction, and is given by
\begin{align}
	\tilde{T}=\frac{\sin\left(\frac{\pi\phi_{2}}{L}\right)\sin\left(\frac{\pi(\phi_3-\phi_4)}{L}\right)}{\sin\left(\frac{\pi\phi_{3}}{L}\right)\sin\left(\frac{\pi(\phi_2-\phi_4)}{L}\right)}\,.\label{T-tilde}
\end{align}

Having determined the Chern-Simons contribution to the BPE, we now move on to the contribution from the Einstein-Hilbert part of the bulk action. From \cref{EE-GMO,BC-disj-gen} the contributions to the entanglement entropy from the Einstein-Hilbert part lead to the constraint equations
\begin{align}
	&\frac{u_{2q}}{\tan\left(\frac{\pi\phi_{2q}}{L}\right)}-\frac{u_{q3}}{\tan\left(\frac{\pi\phi_{q3}}{L}\right)}=\frac{u_{q'2}}{\tan\left(\frac{\pi\phi_{q'2}}{L}\right)}-\frac{u_{q'3}}{\tan\left(\frac{\pi\phi_{q'3}}{L}\right)}\,,\label{EH-BC-GMO1}\\
	&\frac{u_{q'}}{\tan\left(\frac{\pi\phi_{q'}}{L}\right)}-\frac{u_{q'4}}{\tan\left(\frac{\pi\phi_{q'4}}{L}\right)}=\frac{u_{q}}{\tan\left(\frac{\pi\phi_{q}}{L}\right)}-\frac{u_{q4}}{\tan\left(\frac{\pi\phi_{q4}}{L}\right)}\,.\label{EH-BC-GMO2}
\end{align}
The solutions to \cref{EH-BC-GMO1,EH-BC-GMO2} are rather involved and we omit the details here. If we naively substitute these solutions in the Einstein-Hilbert contribution to the PEE $s_{AA'}(A)$, the expression for the BPE still contains the parameters $\phi_q$ and $\phi_{q'}$. Utilizing the solutions obtained in \cref{BC-sol-disj-GMO1}, we finally obtain the Einstein Hilbert contribution to the BPE to be
\begin{align}	
&\frac{12}{c_M}\,\,\text{BPE}^{\text{EH}}(A:B)\notag\\
&=\frac{u_4\left(\sin\frac{2\pi\phi_{23}}{L}+\sin\frac{\pi\phi_3}{L}\right)-u_3\left(\sin\frac{2\pi\phi_{24}}{L}+\sin\frac{\pi\phi_4}{L}\right)+u_{23}\sin\frac{\pi\phi_2}{L}+u_2\left(\sin\frac{2\pi\phi_{34}}{L}-\sin\frac{\pi\phi_3}{L}+\sin\frac{\pi\phi_4}{L}\right)}{4\,\sin\frac{\pi\phi_{23}}{L}\sin\frac{\pi\phi_4}{L} \sqrt{\sin\frac{\pi\phi_2}{L} \sin\frac{\pi\phi_3}{L} \sin\frac{\pi\phi_{24}}{L} \sin\frac{\pi\phi_{34}}{L}}}
\end{align}
Utilizing the cross-ratios for the two disjoint intervals in the  GCFT$_2$ with a finite size, we may rewrite the above expression as
\begin{align}
	\text{BPE}^{\text{EH}}(A:B)=\frac{c_M}{12}\,\left|\frac{\tilde{X}}{\sqrt{\tilde{T}}\left(1-\tilde{T}\right)}\right|\,,
\end{align}
where $\tilde{T}$ is given in \cref{T-tilde} and 
\begin{align}
	\tilde{X}=\tilde{T}\left(\frac{u_2}{\tan\frac{\pi\phi_2}{L}}+\frac{u_{34}}{\tan\frac{\pi\phi_{34}}{L}}-\frac{u_3}{\tan\frac{\pi\phi_3}{L}}+\frac{u_{24}}{\tan\frac{\pi\phi_{24}}{L}}\right)\,.
\end{align}
Therefore the total BPE for the mixed state configuration of two disjoint intervals in a GCFT$_2$ with a finite size dual to the global Minkowski orbifold is given by
\begin{align}
	\text{BPE}(A:B)=\frac{c_L}{12}\log\left(\frac{1+\sqrt{\tilde{T}}}{1-\sqrt{\tilde{T}}}\right)+\frac{c_M}{12}\,\left|\frac{\tilde{X}}{\sqrt{\tilde{T}}\left(1-\tilde{T}\right)}\right|\,.
\end{align} 
Upon using the values of the central charges in \cref{central-charges}, this expression is seen to be exactly equal to the minimal EWCS for the mixed state under consideration \cite{Basu:2021awn}. The above result also matches with the corresponding reflected entropy computed in \cite{Basak:2022cjs,Setare:2022uid} using field theoretic replica techniques.

\subsubsection{Flat space cosmology}
Finally we move on to the computation of the BPE for two disjoint intervals $A=[(u_1,\phi_1),(u_2,\phi_2)]$ and $B=[(u_3,\phi_3),(u_4,\phi_4)]$ in a finite temperature GCFT$_2$  whose bulk dual spacetime is described by TMG in the flat space cosmological geometry. The entanglement entropy for a generic interval $D$ on the asymptotic boundary is given  in \cref{EE-FSC}. As described earlier in the context of the global orbifold, the balance conditions in this case give rise to four constraint equations which are enough to determine the partition points $Q:(u_q,\phi_q)$ and $Q':(u_{q'},\phi_{q'})$. The conditions stemming from the Chern-Simons contributions to the entanglement entropies are given as
\begin{align}
	\frac{\sinh\left(\frac{\pi\phi_{2q}}{\beta}\right)}{\sinh\left(\frac{\pi\phi_{q3}}{\beta}\right)}=\frac{\sinh\left(\frac{\pi\phi_{q'2}}{\beta}\right)}{\sinh\left(\frac{\pi\phi_{q'3}}{\beta}\right)}~~~,~~~\frac{\sinh\left(\frac{\pi\phi_{q'}}{\beta}\right)}{\sinh\left(\frac{\pi\phi_{q'4}}{\beta}\right)}=-\frac{\sinh\left(\frac{\pi\phi_{q}}{\beta}\right)}{\sinh\left(\frac{\pi\phi_{q4}}{\beta}\right)}\label{top-BC-FSC}\,.
\end{align}
As earlier, we set $(u_1,\phi_1)=(0,0)$ for simplicity, utilizing the conformal symmetry of the dual GCFT$_2$. Note that the conformal cross-ratios for this configuration are given by
\begin{align}
	\hat{T}=\frac{\sinh\left(\frac{\pi\phi_{2}}{\beta}\right)\sinh\left(\frac{\pi\phi_{34}}{\beta}\right)}{\sinh\left(\frac{\pi\phi_{3}}{\beta}\right)\sin\left(\frac{\pi\phi_{24}}{\beta}\right)}\,~~,~~
	\hat{X}=\hat{T}\left(\frac{u_2}{\tanh\frac{\pi\phi_2}{\beta}}+\frac{u_{34}}{\tanh\frac{\pi\phi_{34}}{\beta}}-\frac{u_3}{\tanh\frac{\pi\phi_3}{\beta}}+\frac{u_{24}}{\tanh\frac{\pi\phi_{24}}{\beta}}\right)\,.
	\label{cross-FSC}
\end{align}
With this simplification, the solutions to the above constraint equations completely determine the parameters $\phi_q$ and $\phi_{q'}$ as follows
\begin{align}
	&\phi_q=\frac{\beta}{2\pi}\log\left(\frac{2e^{\frac{2\pi\phi_4}{\beta}}-e^{\frac{2\pi\phi_{q'}}{\beta}}-e^{\frac{2\pi(\phi_{4}+\phi_{q'})}{\beta}}}{1+e^{\frac{2\pi\phi_{4}}{\beta}}-2e^{\frac{2\pi\phi_{q'}}{\beta}}}\right)\,,\notag\\
	&\phi_{q'}=\frac{\beta}{2\pi}\log\left(\frac{e^{\frac{2 \pi  (\phi_2+\phi_3)}{\beta }}-4 e^{\frac{\pi  (\phi_2+\phi_3+\phi_4)}{\beta }} \sqrt{\sinh \left(\frac{\pi  \phi_2}{\beta }\right) \sinh \left(\frac{\pi  \phi_3}{\beta }\right) \sinh \left(\frac{\pi  \phi_{24}}{\beta }\right) \sinh \left(\frac{\pi  \phi_{34}}{\beta }\right)}-e^{\frac{2 \pi  \phi_4}{\beta }}}{e^{\frac{2 \pi  \phi_2}{\beta }}+e^{\frac{2 \pi  \phi_3}{\beta }}-e^{\frac{2 \pi  \phi_4}{\beta }}-1}\right)\,.\label{BC-sol-disj1-FSC}
\end{align}
Now utilizing the above solutions to the balance conditions, we may obtain the Chern-Simons contribution to the BPE for the two disjoint intervals as follows
\begin{align}
	\text{BPE}^{\text{CS}}(A:B)&=\frac{c_L}{12}\log\left[\frac{\sinh\left(\frac{\pi\phi_{1q}}{\beta}\right)\sinh\left(\frac{\pi\phi_{q'2}}{\beta}\right)}{\sinh\left(\frac{\pi\phi_{2q}}{\beta}\right)\sinh\left(\frac{\pi\phi_{q'1}}{\beta}\right)}\right]\notag\\&=\frac{c_L}{12}\log\left(\frac{1-\sqrt{\hat{T}}}{1+\sqrt{\hat{T}}}\right)\,,
\end{align}
where $\hat{T}$ is given in \cref{cross-FSC}.

Next we compute the Einstein-Hilbert contribution to the BPE for the two disjoint intervals in the thermal GCFT$_2$. The balance conditions in this case are given by
\begin{align}
		&\frac{u_{2q}}{\tanh\left(\frac{\pi\phi_{2q}}{\beta}\right)}-\frac{u_{q3}}{\tanh\left(\frac{\pi\phi_{q3}}{\beta}\right)}=\frac{u_{q'2}}{\tanh\left(\frac{\pi\phi_{q'2}}{\beta}\right)}-\frac{u_{q'3}}{\tanh\left(\frac{\pi\phi_{q'3}}{\beta}\right)}\,,\label{EH-BC-FSC1}\\
	&\frac{u_{q'}}{\tanh\left(\frac{\pi\phi_{q'}}{\beta}\right)}-\frac{u_{q'4}}{\tanh\left(\frac{\pi\phi_{q'4}}{\beta}\right)}=\frac{u_{q}}{\tanh\left(\frac{\pi\phi_{q}}{\beta}\right)}-\frac{u_{q4}}{\tanh\left(\frac{\pi\phi_{q4}}{\beta}\right)}\,.\label{EH-BC-FSC2}
\end{align}
Substituting the solutions to the above constraints along with those in \cref{BC-sol-disj1-FSC} in the expression for the partial entanglement entropy $s_{AA'}(A)$ given in \cref{PEE-disj-PM}, we may obtain the Einstein-Hilbert contribution to the BPE for the present configuration. The result may be summarized in terms of the cross-ratios given in \cref{cross-FSC}. The final expression for the BPE for the mixed state configurtion described by two disjoint intervals in the thermal GCFT$_2$ located at the null infinity of the flat space cosmological spacetime is then given by
\begin{align}
	\text{BPE}(A:B)=\frac{c_L}{12}\log\left(\frac{1+\sqrt{\hat{T}}}{1-\sqrt{\hat{T}}}\right)+\frac{c_M}{12}\left|\frac{\hat{X}}{\sqrt{\hat{T}}\left(1-\hat{T}\right)}\right|\,.
\end{align}
Once again, the above BPE is exactly equal to the minimal EWCS as well as half of the reflected entropy in \cite{Basak:2022cjs,Setare:2022uid} for the present mixed state configuration of two disjoint intervals in the thermal GCFT$_2$. This serves as yet another consistency check of our holographic construction.

\subsection{Geometric picture for BPE and the EWCS}
In this section we provide a geometric notion of the BPE obtained earlier, utlizing the novel geometric picture of the holographic entanglement entropy devised in \cite{Jiang:2017ecm,Wen:2018mev} in the context of flat space holography. We will also discuss the natural extension of the entanglement contour or the partial entanglement entropy studied in \cite{Wen:2018whg,Han:2019scu,Wen:2019iyq,Wen:2020ech} to the setup of flat holography as described in \cite{Wen:2018mev,Wen:2020ech}. Furthermore, we will utilize the construction of the entanglement wedge described in \cite{Basu:2021awn} and briefly alluded to in subsection \ref{sec:flat-holography}, to demonstrate the equivalence of the BPE and the minimal EWCS in such flat holographic settings.

In \cite{Wen:2018whg}, a novel fine-grained holographic picture of the entanglement contour function was put forward based on the equivalence between the bulk and boundary modular flows in the usual AdS/CFT scenario. Essentially the fine correspondence dictates that the contribution to the entanglement entropy for a subsystem $A$ from one of its subsets $A_i$ is given by the length of the chord of the RT surface of $A$, cut out by normal geodesics emanating from the endpoints of $A_i$. Such fine correspondence was shown to be valid in holographic setups beyond AdS/CFT in \cite{Wen:2018mev}. In the case of flat holography, we may expect a similar fine structure correspondence between the PEE and segments of the HRT like extremal curves described in \cite{Jiang:2017ecm,Hijano:2017eii}. One way to convince ourselves is to consider the PEE in GCFT$_2$s dual to asymptotically flat spacetimes as the flat limit of the AdS/CFT results by taking the AdS radius to infinity.

\begin{figure}[ht]
	\centering
	\includegraphics[scale=0.8]{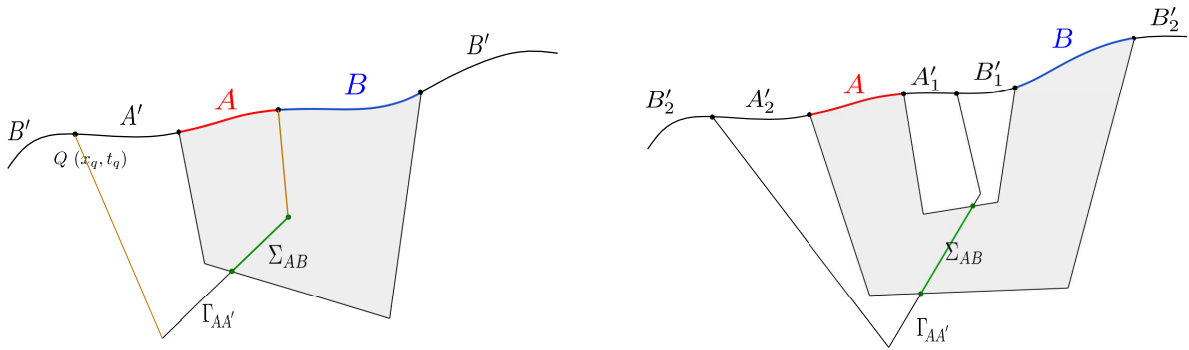}
	\caption{Schematics of the BPE in terms of the partial entanglement entropy and the minimal EWCS. }
	\label{fig:geom}
\end{figure}

Similar to the AdS/CFT scenario, the minimal EWCS $\Sigma_{AB}$ in flat holographic setups is an extremal curve passing through the HRT like surface $\Gamma_{AB}$ computing the entanglement entropy of the total system $A\cup B$ \cite{Basu:2021awn}. Following the prescription in \cite{Wen:2021qgx} for the usual AdS/CFT scenario, we may extend the extremal segment $\Sigma_{AB}$ up to the boundary. To do so, we need to introduce two additional null geodesics starting from the endpoints of the extended segment of the extremal curve to two different points on the boundary. This procedure essentially extends the EWCS $\Sigma_{AB}$ to the HRT surface of some specific subsystem on the boundary. As an example, consider the case of two adjacent intervals $A$ and $B$ shown in the left panel in \cref{fig:geom}. In this case, one end of the EWCS $\Sigma_{AB}$ lands on the boundary through a null geodesic. We extend $\Sigma_{AB}$ further into the bulk and then through another null geodesic landing on some point $P$ on the boundary, which partitions the complement of $A\cup B$ into two parts $A'$ and $B'$, as shown in \cref{fig:geom}. Now using the geometric picture of the entanglement entropy in \cite{Jiang:2017ecm,Hijano:2017eii}, this extended extremal curve is nothing but the HRT surface for the subsystem $A\cup A'$ (or equivalently, the HRT surface for $B\cup B'$). If we now assume that the  geometric construction for the fine structure correspondence between the PEE and geodesic chords described in \cite{Wen:2018whg,Wen:2021qgx} has a well defined flat limit, then it is evident that the area of $\Sigma_{AB}$ plays the role of the PEE $s_{AA'}(A)$ (or equivalently $s_{BB'}(B)$),
\begin{align}
	s_{AA'}(A)=\frac{\text{Area}\left(\Sigma_{AB}\right)}{4G_N}=s_{BB'}(B)\,.
\end{align}
Therefore, the balance condition is naturally satisfied by the partition induced by the extension of $\Sigma_{AB}$ leading to the equivalence of the BPE and the area of the minimal EWCS in flat holography,
\begin{align}
	\text{BPE}(A:B)=\frac{\text{Area}\left(\Sigma_{AB}\right)}{4G_N}\,.
\end{align}
We may also establish the above equality for the case of two disjoint intervals sketched in the right panel of \cref{fig:geom} using similar arguments.
\section{Summary and conclusions}
\label{sec:summary}
To summarize, in this article we have obtained the balanced partial entanglement entropy for various bipartite mixed state configurations in a class of $(1+1)$-dimensional non-relativistic field theories with Galilean conformal symmetry which are dual to Einstein gravity and topologically massive gravity in asymptotically flat $3d$ spacetimes. To this end, we have devised a method to obtain the minimal balanced purification by introducing suitable bipartitions of the auxiliary purifier subsystems in such flat holographic settings where observables depend on the specific choice of the frame. We investigate the BPE for mixed state configurations involving two adjacent intervals in the ground state of a GCFT$_2$, for a GCFT$_2$ with a finite size as well as in a thermal GCFT$_2$. Interestingly, our results match exactly with the minimal EWCS for each of the mixed state configurations for dual  spacetime geometries described by pure Einstein gravity as well as TMG. Furthermore, the recent results for the reflected entropy in such GCFT$_2$s agree with our findings for the BPE in the limit of large central charges.

We have also investigated the crossing correlation between original and purifier subsystems in terms of the crossing PEEs for the mixed state configurations described by two adjacent intervals.
In particular, we have found that for mixed states involving two adjacent intervals in GCFT$_2$s dual to pure Einstein gravity in the bulk, the crossing correlations at the balance point vanish. As described in the literature, such crossing PEEs are the natural generalization of the so called Markov gap which characterizes the precision of a Markov recovery process. Therefore, the vanishing crossing PEE indicates the possibility of a perfect Markov recovery in flat holographic scenarios involving pure Einstein gravity in the bulk. Remarkably, for TMG in the bulk dual geometries, the crossing PEE turns out to be a universal constant and has a topological origin. Furthermore, this universal crossing PEE is reminiscent of the geometric interpretation of the Markov gap in the usual AdS/CFT settings in terms of the number of boundaries of the EWCS.

Subsequently, we have obtained the BPE for bipartite mixed states involving two disjoint intervals in such holographic GCFT$_2$s. Recently in the context of holographic BMS field theories in $(1+1)$-dimensions dual to Einstein gravity without a cosmological constant, it was found that for covariant configurations involving non-adjacent intervals, one requires further auxiliary constraints along with the balance conditions. In this work we have found that such additional constraints naturally stem from the topological Chern-Simons contributions to the PEEs when the full topologically massive gravity theory is considered. Remarkably, the BPE for the mixed state of two disjoint intervals match exactly with the corresponding area of the minimal EWCS reported in earlier literature.
Furthermore, the large central charge behavior of the reflected entropy obtained through field theoretic replica techniques also agrees with our findings for the BPE in the mixed states under consideration. 

There are several interesting future directions to investigate. One workable case is the holographic correspondence between CFTs with gravitational anomaly and TMG in asymptotically AdS spacetimes. The holographic entanglement entropy for such theories was investigated in \cite{Castro_2014}. Furthermore, the reflected entropy and the EWCS for various bipartite states were computed recently in \cite{Basu:2022nds}. It will be interesting to study the correspondence of BPE with the EWCS or the reflected entropy in these anomalous theories. The geometric picture for the entanglement structure in higher dimensional flat holographic scenarios was investigated in \cite{Godet:2019wje}. The EWCS in such higher dimensional settings may be constructed in a similar way to \cite{Basu:2021awn}. It will also be interesting to test the duality between the BPE and the EWCS in these setups. We hope to return to these issues in future.

	\bibliographystyle{utphys}
	\bibliography{reference}
	
\end{document}